\documentclass[%
 reprint,
 amsmath,amssymb,
 aps,
]{revtex4-2}

\usepackage[normalem]{ulem}
\usepackage{graphicx}% Include figure files
\usepackage{dcolumn}% Align table columns on decimal point
\usepackage{bm}% bold math
\usepackage{hyperref}% add hypertext capabilities
\usepackage{soul,xcolor}
\setstcolor{red}

\begin{document}

\title{Relaxation shortcuts through boundary coupling}% Force line breaks with \\

\author{Gianluca Teza}
\author{Ran Yaacoby}%
\author{Oren Raz}
 \email{oren.raz@weizmann.ac.il}
\affiliation{%
 Department of Physics of Complex Systems, Weizmann Institute of Science, Rehovot 7610001, Israel
}%

\date{\today}

\begin{abstract}
When a hot system cools down faster than an equivalent cold one, it exhibits the Mpemba Effect. This counterintuitive phenomenon was observed in several systems including water, magnetic alloys and polymers. In most experiments the system is coupled to the bath through its boundaries, but all theories so far assumed bulk coupling.
Here we build a general framework for boundary coupling relaxation and show that the Mpemba effect persists in these cases. Surprisingly, it can survive even an arbitrarily weak couplings. An example is given in the Ising antiferromagnetic chain.
\end{abstract}

\maketitle

When coupled to a thermal bath, most systems relax towards equilibrium. The precise details of the relaxation are determined by many factors, including the intrinsic properties of the specific system, its initial condition, the bath's properties and the exact nature of the coupling between the system and the bath. However, it is commonly expected that in the weak coupling limit a macroscopic system that is initiated in equilibrium state corresponding to some initial temperature $T_0$ relaxes \emph{quasi-statically} towards the bath temperature $T_b$, such that the system can be described at each instance as in equilibrium for some temperature. This is a consequence of the self-thermalization being much faster then the heat exchange with the thermal bath. In a strong coupling, on the other hand, the self-thermalization process through which the system equilibrates is not fast enough, and the energy exchange with the environment drives the system into a relaxation trajectory that can reach far from any equilibrium distributions where anomalous relaxations might arise. Such far from equilibrium relaxation trajectories can be counterintuitive, and show interesting phenomena that are unexpected near equilibrium \cite{lapolla2020faster,Gal2020,Lu2017}. An important example for such a phenomenon is the Mpemba Effect (ME) \cite{Aristotele350, Mpemba1969}, where a hot system, under proper conditions, cools down faster than an initially cold one when quenching both to an even colder bath. The ME was observed experimentally in a variety of setups, including water \cite{Jeng2006}, magnetic alloys \cite{Chaddah2010}, polymers \cite{Hu2018}, clathrate hydrates \cite{paper:hydrates} and very recently in small size systems like colloids diffusing in a potential \cite{Kumar2020,kumar2021anomalous}. In addition to these experimental observations, it was also observed in a variety of numerical and theoretical models for water molecules \cite{Tao2016,Zhang2014,Vynnycky2015,Auerbach1998,Mirabedin2017}, driven granular gases \cite{Lasanta2017,torrente2019large,biswas2020mpemba,santos2020mpemba}, inertial suspensions \cite{takada2021mpemba}, gas of visco-elastic particles \cite{mompo2021memory}, diffusing in a potential \cite{chetrite2021metastable,Gal2020,walker2021anomalous,busiello2021inducing}  and classical as well as quantum spin models \cite{Carollo2021,baity2019mpemba,nava2019lindblad,yang2020non,vadakkayil2021should,Klich2019}.

%The framework of Markovian dynamics provides a comprehensive and general characterization of the ME \cite{Lu2017, Klich2019, Gal2020}, a crucial requirement when trying to model a phenomenon spanning across such an heterogeneous variety of systems.
%Many physical and chemical processes can indeed be ascribed to a Markovian dynamics, allowing one to fully characterize the relaxation process in terms of a linear evolution operator \cite{Kampen2007}.
%This framework provides a clear mechanism, some physical intuition and even new predictions -- the inverse \cite{Lu2017} and strong \cite{Klich2019} effects that have been recently observed experimentally \cite{Kumar2020,kumar2021anomalous}. 
The theoretical models proposed so far to explain the ME assumed that all the relevant degrees of freedom (e.g. all spins, or all molecules) are directly coupled to the thermal bath. However, in most experiments demonstrating the ME (with colloidal system \cite{Kumar2020,kumar2021anomalous} being the only exception), the system is coupled to the heat bath only through its boundaries, potentially hindering the adequacy of such models for large sized systems. Similarly, the dependence of the effect on the coupling strength is unknown. Intuitively, one expects the effect to become visible as the coupling with the thermal bath strengthens, while a weak coupling, on the other hand, is expected to make the effect harder to observe.

In this letter we construct a general theoretical framework for boundary coupling with the bath, and use it to demonstrate the existence of the ME even in such systems. We do so by combining two types of dynamics: heat exchange through the boundaries and an energy conserving self-thermalization dynamic for the thermally isolated bulk. The ratio between the characteristic timescales of the two dynamics tunes the coupling strength, and allows us to explore the limiting cases of arbitrary weak and strong couplings. We then demonstrate this general construction in the 1D Ising antiferromagnet, and show that it exhibits a ME even at arbitrarily  weak boundary coupling with the bath. An exact coarse-graining of the system \cite{Teza2020,Teza2020b} enables exploring the infinite and zero coupling strength limits, and somewhat surprisingly shows that the effect survives the arbitrarily weak coupling limit.
%Finally, we show how the observation of the ME is strictly connected with another anomalous relaxation phenomenon known as \emph{eigenvalue crossing}, in which the system exhibits a change in the hierarchy of the eigenvalues regulating the relaxation dynamics with respect to some model parameter.

\begin{figure}
    \centering
    \includegraphics[width=\linewidth]{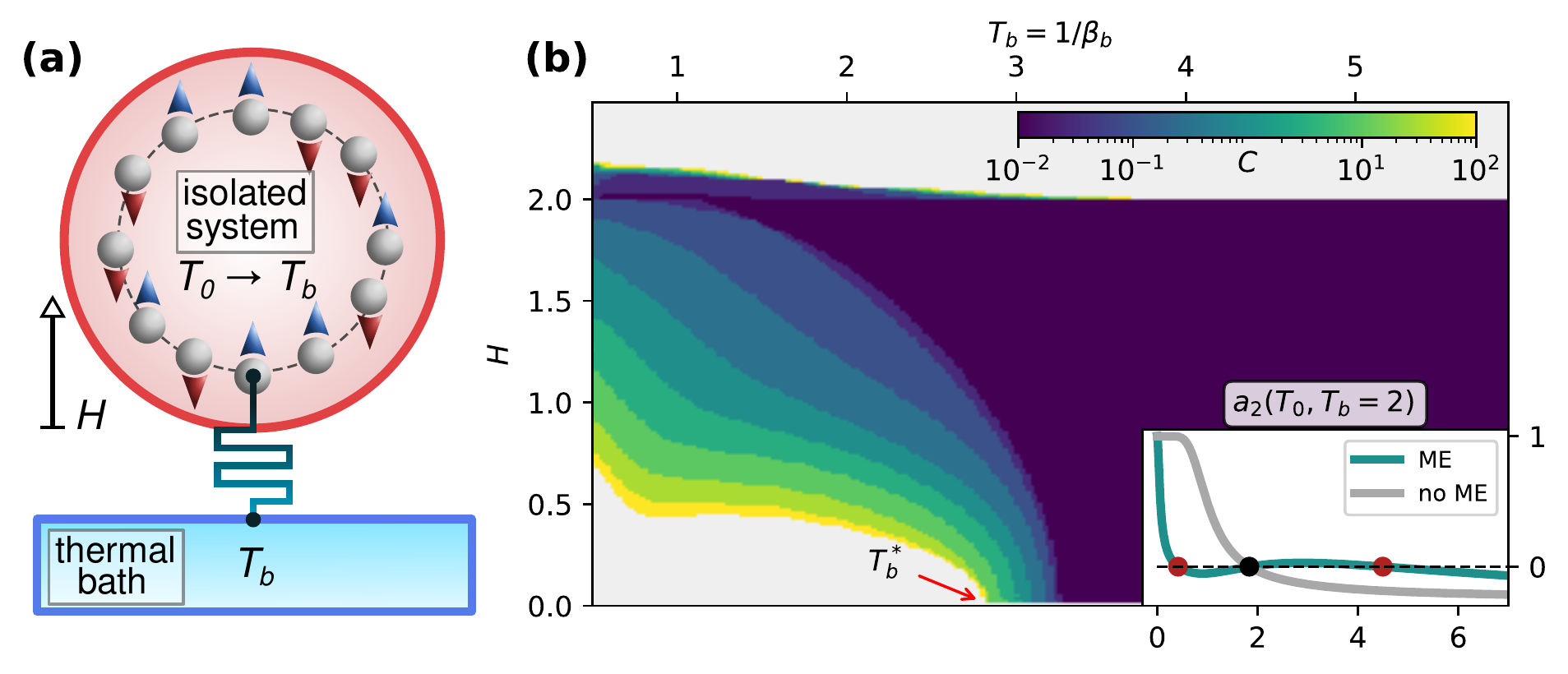}
    \caption{(a) The system: an antiferromagnet Ising chain, with a single spin coupled to the thermal bath. Transitions are allowed only if energy is conserved, except for flipping the spin coupled to the bath. (b) Minimal coupling strength C (colorbar) for which there exists some type of a ME at the corresponding bath temperature $T_b$ (x axis) and $H$ (y axis). In the white areas there is no ME at any coupling strength. Inset: the ME exists if $a_2(T_0,T_b)$ is non-monotonic in $T_0$.
    }
    \label{fig:ph_diag}
\end{figure}

We begin by introducing a general formulation of the setup. To model boundary coupling with the thermal environment, which is stochastic in nature, we consider the probability distribution $\vec{p}$ where the component $p_i(t)$ is the probability to be in a microstate $i$, that has energy $E_i$, at a given time $t$ \footnote{Similar analysis can be made in continuous frameworks as in \cite{Lu2017}}. $\vec{p}(t)$ evolves in time according to the master equation
\begin{equation}\label{eq:mast_eq}
    \partial_t \vec{p}(t)=\mathbf{W}(\beta_b) \vec{p}(t),
\end{equation}
where the rate matrix $\mathbf{W}$ encodes the specific model and depends on the bath (inverse) temperature $\beta_b=1/T_b$ (for simplicity we use units where $k_B=1$). The off-diagonal terms $\mathbf{W}_{ij}$ are the jumping rates from state $j$ to state $i$, while the diagonal term $\mathbf{W}_{ii}=-\sum_{j\neq i}\mathbf{W}_{ji}$ represents the escape rate from the state $i$.
We assume that detailed balance and ergodicity hold, so that regardless of the initial condition the system relaxes towards the (unique) Boltzmann distribution $\pi_i(\beta_b) =e^{-\beta_b E_i}/\mathcal{Z}(\beta_b)$ where $\mathcal{Z}(\beta_b)=\sum_i e^{-\beta_b E_i}$ is the partition function of the system at the bath temperature.

To characterize the relaxation process one formally integrates Eq. \ref{eq:mast_eq}.
Starting from equilibrium conditions at an initial temperature $\beta_0$, we get:
\begin{align}\label{eq:mast_eq_solution}
    \vec{p}(t,\beta_0,\beta_b) &=e^{\mathbf{W}(\beta_b)t}\vec{\pi}(\beta_0) = \\ 
    &=\vec\pi(\beta_b) + \sum_{i>1} a_i(\beta_0,\beta_b) e^{\lambda_i(\beta_b) t} \vec{v}_i(\beta_b), \nonumber
\end{align}
where $0=\lambda_1>\lambda_2\geq\lambda_3\geq\dots$ and $\vec{v}_i$ are respectively the (real) eigenvalues and right eigenvectors of the rate matrix $\mathbf{W}$, while the coefficients $a_i$ are the projection of the left eigenvectors of $\mathbf{W}$ over the initial equilibrium distribution $\vec{\pi}(\beta_0)$. The existence of the ME is encoded in the slowest relaxation, regulated by the second dominant eigenvalue $\lambda_2$ \cite{Lu2017}: a non-monotonic dependence of the coefficient $a_2(\beta_0,\beta_b)$ with respect to the initial temperature $\beta_0$ guarantees the possibility of observing the effect when quenching the system to that specific bath temperature $\beta_b$. For simplicity, in this work we do not distinguish between the different types of the Mpemba effects (inverse vs. direct, strong vs. weak, etc') \cite{Klich2019}.

In many experimental setups, the thermal bath is not directly coupled to all the degrees of freedom composing the system: heat can be transferred only through those sitting on the boundaries.
All other transitions between microstates are ``bulk transitions", and as no energy is exchanged with the bath in such transitions they can only happen between states with the same energy. These bulk transitions serve as a \emph{self-thermalization} (ST) mechanism, whereas the \emph{boundary transitions} (BT) couple the system to the bath and enable transitions between different energy shells.
This structure can be modeled by
\begin{equation}\label{eq:full_rate_matrix}
    \mathbf{W}(\Gamma^{ST},\Gamma^{BC})=\Gamma^{ST}\mathbf{W}^{ST}+\Gamma^{BC}\mathbf{W}^{BC}.
\end{equation}
Here $\mathbf{W}^{ST}$ and $\mathbf{W}^{BC}$ are normalized rate matrices corresponding to the self thermalization and boundary coupling transitions respectively, and $\Gamma^{\{ST,BC\}}$ are coupling constants modulating the rates amplitude. Their ratio, $C=\Gamma^{BC}/\Gamma^{ST}$, dictates the coupling strength \footnote{The specific normalization chosen for $\mathbf{W}^{\{BC,ST\}}$ changes the value of $C$, but not its limiting cases. Specifically, we chose to normalize with respect to the maximum rate so that $\max(\mathbf{W}_{ij}^{ST})=\max(\mathbf{W}_{ij}^{BC})\equiv 1$.}: in the limit $C\ll 1$ boundary flips -- regulating the energy exchange with the bath -- occur rarely compared to thermalization flips, so the system thermalizes quickly after each energy exchange with the bath. In the $C\gg1$ limit, the boundaries (in thermal equilibrium with the bath) exchange heat much faster than the thermalization and the diffusion of energy within the system sets the timescale for the relaxation. We refer to the former limit as the \emph{weak coupling} and to the latter as \emph{strong coupling}. 

By construction, $\mathbf{W}^{ST}$ is a rate matrix that contains only transitions between states that have the same energy. Generically, it is a reducible matrix with a zero eigenvalue degeneracy equal to the number of different energy shells. As for $\mathbf{W}^{BC}$, it is generically a sparse matrix, as most transitions involve more than the boundary spins. The interaction of the system with a thermal bath at inverse temperature $\beta_b$ satisfies detailed balance, therefore $\mathbf{W}^{BC}_{ji} = \mathbf{W}^{BC}_{ij}e^{-\beta_b (E_j-E_i)}$, where $E_{\{i,j\}}$ are the energy corresponding to the two states.
%\textcolor{red}{we already mentioned DB before, maybe we should put this there. Also, the conditions holds trivially for WST.}

In the weak coupling limit, a naive perturbation scheme with $C\ll1$ would not prove useful: for $C=0$ the matrix $\mathbf{W}$ is reducible and its zero eigenvalue is highly degenerate, so one cannot apply the standard analysis. Instead, it is constructive in this case to aggregate all the microstates that share the same energy into a single \emph{macrostate} and construct the effective dynamics by summing all the microscopic transitions between them \cite{Teza2020,Teza2020b}. 
In this case, the dynamics is dictated only by the boundary flips, and the diffusion within each energy shell is assumed to happen instantaneously. 
Similarly, in the strong coupling limit, micro-states can be aggregated into macrostates by combining all the microstates connected by boundary flips. Mathematically, the two aggregation procedures can be done by arranging the states such that $\mathbf{W}^{BC}$ or $\mathbf{W}^{ST}$ is block diagonal where each block corresponds to transitions within a macrostate, and coarsening over these blocks.

Let us demonstrate the above construction with a specific example, depicted in the cartoon of Fig. \ref{fig:ph_diag}(a). It consists in a 1-dimensional ring of $N$ Ising spins with nearest neighbour antiferromagnet interactions. Each spin $\{\sigma_s\}_{s=1\dots N}$ in the chain can either be in an up or down ($\pm1$) state, giving a total of $M=2^N$ different microstates, identified by the $N$-dimensional vector $\vec{\sigma}$. The energy of a microstate $\vec{\sigma}$ is provided by the Hamiltonian functional
\begin{equation}\label{eq:ham_AF}
    \mathcal{H}(\vec{\sigma})= - J\sum_{ s } \sigma_{s} \sigma_{s+1} -H\sum_{s=1}^{N} \sigma_{s},
\end{equation}
where $J<0$ is the coupling constant, $H$ is an external magnetic field and $\sigma_{N+1}\equiv \sigma_1$. For simplicity, we set $J=-1$.

To model boundary coupling in this system, we choose a specific spin (say $\sigma_1$), which is coupled to the bath. This implies that a general microstate $\vec\sigma$ is connected through thermal flips only with a single state $\vec\sigma'$ in which the first spin is flipped, $\sigma_{1}\to -\sigma_{1}$, while the remaining spins are unaltered. For two general microstates $\vec \sigma^i$ and $\vec \sigma^j$ the transition is therefore
\begin{equation}\label{eq:W_BC}
    \mathbf{W}^{BC}_{ij}= \frac{\delta_{\sigma_{1}^i,-\sigma_{1}^j}\prod_{s>1}\delta_{\sigma_{s}^{i},\sigma_{s}^j}}{1+e^{\beta_b \left(\mathcal{H}\left(\vec{\sigma}^i\right)-\mathcal{H}\left(\vec{\sigma}^j\right)\right)}}
\end{equation}
where $\delta_{ij}$ is the Kronecker delta, $\sigma_s^i$ is the $s$ spin in the microstate $\vec{\sigma}^i$ and we used standard Glauber dynamics as the transition weight \cite{Glauber1963,Felderhof1971}, ensuring that the equilibrium distribution is the Boltzmann distribution.

To model bulk transitions we use rates that decay exponentially as $2^{-d_{ij}}$, where $d_{ij}=\sum_s \delta_{\sigma_{s}^i,-\sigma_{s}^j}$ is the Hamming distance \cite{Hamming1950} that counts the number of spins that has to be flipped between the two configurations, as suggested by decimation-like procedures operated on Markov jump processes \cite{Teza2020,Teza2020b}. We therefore formalize bulk transitions between two states $i\neq j$ as
\begin{equation}\label{eq:W_ST}
    \mathbf{W}^{ST}_{ij}= \delta_{\mathcal{H}\left(\vec{\sigma}^i\right),\mathcal{H}\left(\vec{\sigma}^j\right)}2^{-d_{ij}}.
\end{equation}
The full transition matrix for the model is finally built as a linear combination of the two rate matrices as in Eq. \ref{eq:full_rate_matrix}.

With this construction, let us consider the persistence of the ME in a boundary coupling setup.
In Fig. \ref{fig:ph_diag}\textbf{(b)} we plot the minimal coupling constant $C$ for which some type of a ME exists in the system, for $N=10$ (implying $M=1024$ microstates) and as a function of bath temperature $\beta_b$ and external magnetic field $H$.
The strength of the coupling affects only \emph{quantitatively} the regions where the ME can be observed (the larger $C$, the larger the area). In particular, we see that for any $|H|\leq2$ the effect exists for any bath temperature above a critical $T_b^*=1/\beta^*$ (highlighted with a red arrow).

\begin{figure}
    \centering
    \includegraphics[width=\linewidth]{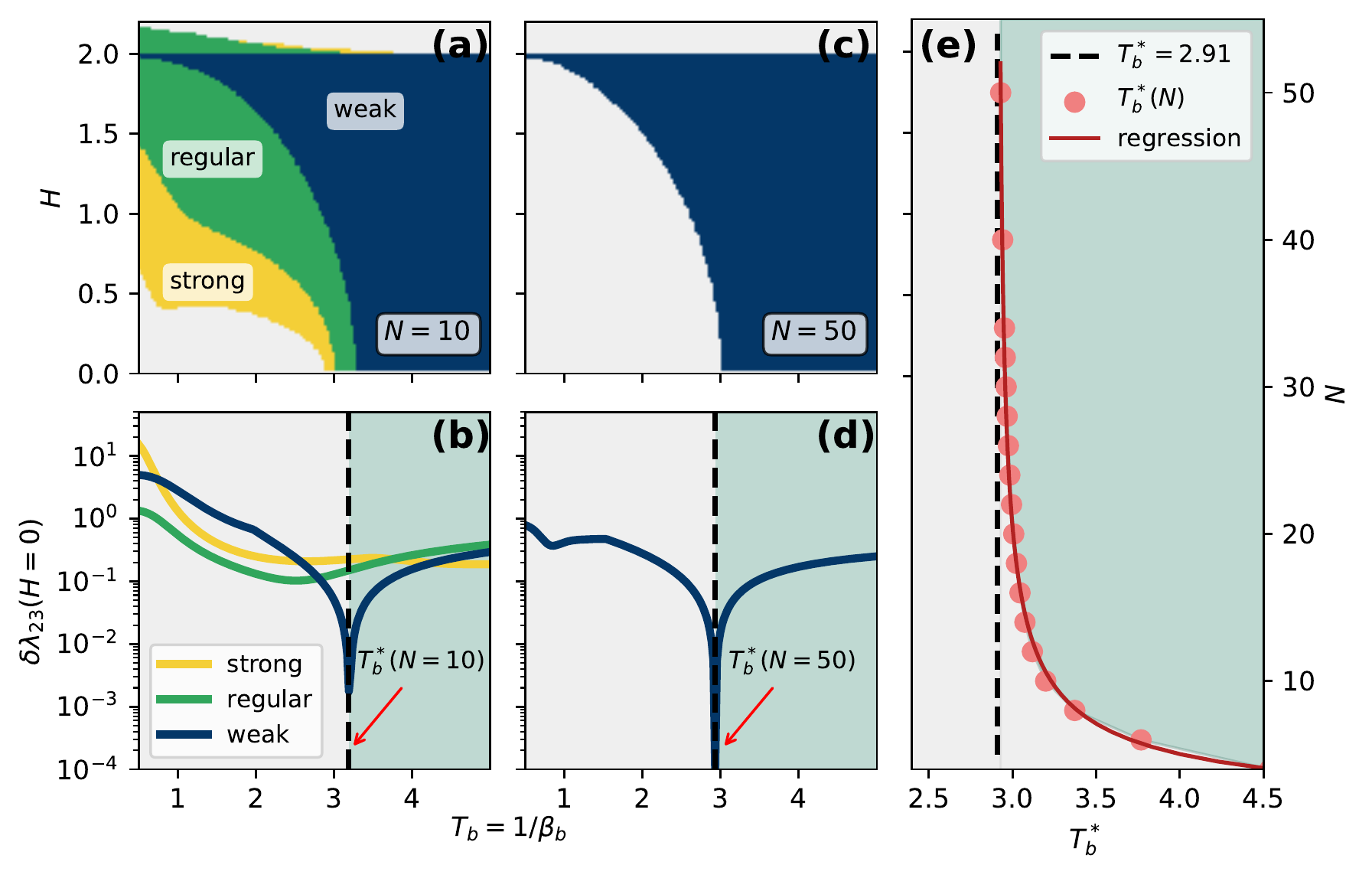}
    \caption{\textbf{(a)} {Comparison of the ME for different coupling strengths ($N=10$). The strong coupling limit ($C\gg 1$) includes all colored areas, while the intermediate coupling ($C=1$) is limited to the green and blue ones. Surprisingly, the effect survives the arbitrarily weak  coupling limit ($C\ll 1$, blue area).}
    \textbf{(b)} Distance between the second and third eigenvalues as a function of $T_b$ for $H=0$. A crossing of the eigenvalues clearly marks the beginning of the Mpemba region at $T_b^*$ for the weakly coupled model.
    \textbf{(c,d)} Phase diagram in the weak coupling limit for $N=50$.
    \textbf{(e)} $T_b^*$ as a function of $N$. The critical value is extracted, through polynomial feeting, to be $T_b^*\sim2.91$}
    \label{fig:ph_diag_ws}
\end{figure}

In the weak coupling case {($C\ll1$)}, the coarse-grained rate matrix is given by 
\begin{equation}
    \mathbf{W}^{\text{weak}}_{ij}=\frac{G_{ij}}{\Omega_j}\frac{1}{1+e^{\beta_b(E_i-E_j)}}
\end{equation}
where the indices $i,j$ now refer to the energies $E_i$ and $E_j$, the (symmetric) matrix $G_{ij}$ counts the number of transitions connecting microstates in the two energy shells, and $\Omega_i$ is the number of microstates with energy $E_i$. This coarsening considerably reduces the size of the matrix, allowing to numerically analyze longer chains assessing the stability of the phase diagram in the thermodynamic limit. The total number of energy shells grows quadratically as $2+(N/2)^2$, as opposed to the exponential growth of the number of microstates. As an example, at $N=50$  (Fig. \ref{fig:ph_diag_ws}c,d) there are $\sim10^{15}$ microstates, but only  627 macrostates in the coarse-grain representation.

{The case of extremely strong coupling ($C\gg1$) can be similarly analyzed. In our model only a single spin is coupled to the bath, therefore the clustering of a $N$ spins chain model results in an effective $N-1$ long chain with an additional ``superposed'' spin oscillating infinitely fast between the two $\pm1$ states. Indicating with $i$ one of the possible $2^{N-1}$ configurations of the bulk chain, we set $\mathcal{H}_i^{\pm1}$ to be the Hamiltonians of each of the two possible states in the $i$-th cluster. The two states composing a cluster are not equivalent as in the weak coupling case. To correctly define the transition rates in the coarse-grained model we therefore need to introduce a Glauber weight: $w^{\sigma}_j=e^{-\beta_b\mathcal{H}^\sigma_j}/(e^{-\beta_b\mathcal{H}^{-1}_j}+e^{-\beta_b\mathcal{H}^{+1}_j})$ with $\sigma=\pm1$ depending on the microstate from which the original transition occurred.
This provides us with:
\begin{equation}
    \mathbf{W}^{\text{strong}}_{ij}=\sum_{\sigma,\sigma'} w^{\sigma}_j \delta_{\mathcal{H}^\sigma_j,\mathcal{H}^{\sigma'}_i}
    2^{-(d_{ij}+\delta_{\sigma,-\sigma'})}
\end{equation}
where the Kroneker delta corrects the Hamming distance for the coupled spin.
The area in which an effect can be observed is wider (Fig. \ref{fig:ph_diag_ws}a): a stronger coupling should indeed ease the undertake of anomalous relaxation paths.}
%However, in some regions the effect is still not observed, implying that the ME cannot just be obtained imposing arbitrarily strong couplings, another strong point towards the fact that both boundaries and coupling strength alone can't determine the sole existence of the ME.

In Fig. \ref{fig:ph_diag_ws}(a) we plot the regions in which some ME can be observed in the limiting coupling setups discussed above, and compare them with  the intermediate $C=1$ case. Somewhat surprisingly, {the} ME can be observed even for $C\ll 1$, demonstrating that the effect survives the limit of arbitrarily weak coupling.

\begin{figure}
    \centering
    \includegraphics[width=\linewidth]{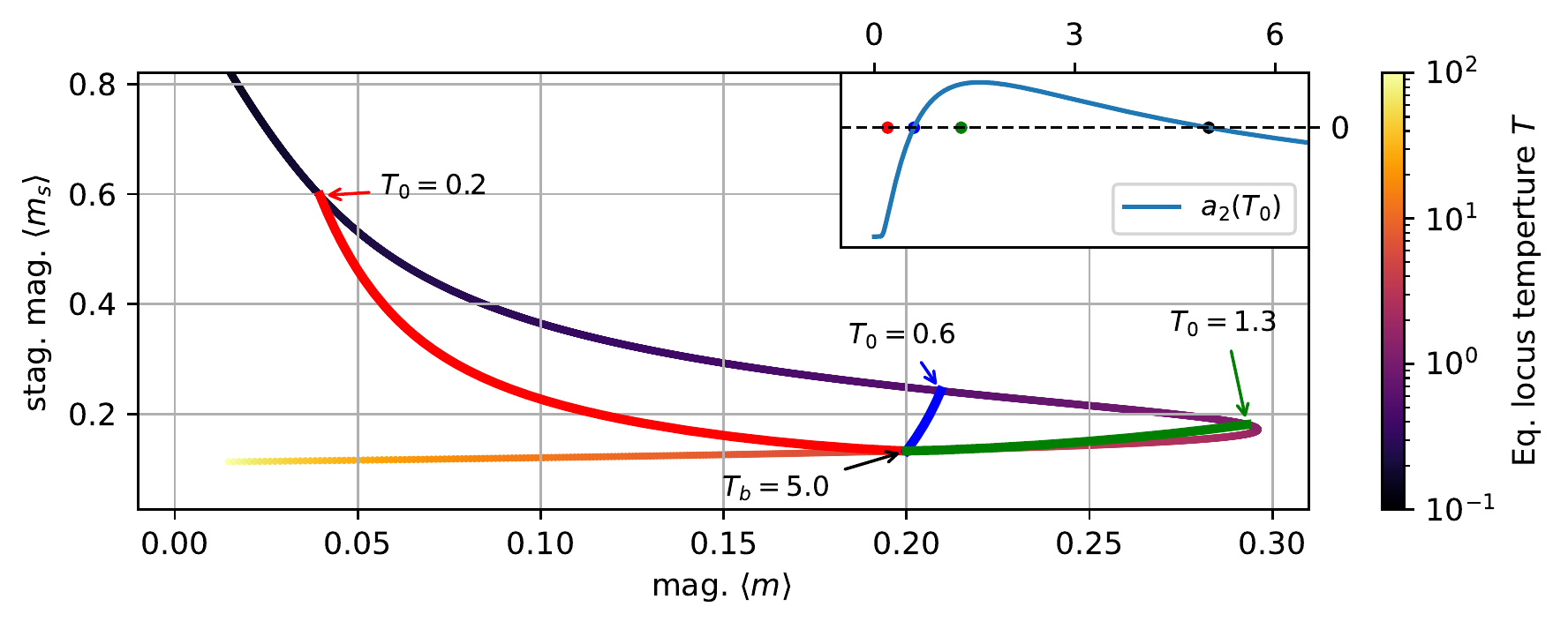}
    \caption{Relaxation paths for a weakly coupled system ($N=50$) on the $\left< m \right>$-$\left< m_s \right>$ plane.
    Initial temperatures are chosen before (red), at (blue) and after (green) the zero of $a_2(T_0)$ as shown in the inset.
    Red and green paths approach the equilibrium locus along the same vector with opposite direction, indicating the existence of a ME.}
    \label{fig:relax_path}
\end{figure}

Additional analysis that can demonstrate the ME in this setup {can be obtained by projecting} the relaxation paths in the (high-dimensional) probability space on two order parameters: the average $\left< m \right>$ and staggered magnetization $\left< m_s \right>$ \cite{Gal2020}. This method enables to use Monte-Carlo simulation in large systems where direct calculation of $a_2$ is impractical, and nevertheless compare the relaxation trajectory with the equilibrium distributions at different temperatures. In Fig. \ref{fig:relax_path} we illustrate different relaxation paths in a weakly coupled chain of $N=50$ spins, in comparison to the equilibrium line.
Initial temperatures are chosen before, at and after the zero of $a_2$ at $T_0\sim0.6<T_b$. The hot (green) and cold (red) relaxation paths approach the bath equilibrium at $T_b=5$  along  the  same  direction, which is the projection of $\vec{v}_2$ into the observable space, but with opposite direction. This {corresponds to} a sign-change in $a_2$, which indeed vanishes at the blue trajectory {implying non-monotonicity of $a_2(T_0)$ (inset of Fig. \ref{fig:relax_path}) and consequently a ME in the system.}

An important feature emerging from the weak coupling limit analysis links the ME with a degeneracy in the spectrum of the rate matrix $\mathbf{W}$.
The ME is defined through the monotonicity of $a_2(\beta_0,\beta_b,H)$ as a function of $\beta_0$, which in general is expected to change continuously. However, at $H=0$ and a specific bath temperature which we denote by $T_b^*$, the first non-zero eigenvalues overlap: $\lambda_2=\lambda_3$  (See Fig. \ref{fig:ph_diag_ws}b,d).
At this point, $\vec{v}_2$ and $\vec{v}_3$ exchange their role as the slowest direction in the system, and consequently $a_2$ changes discontinuously and the ME appears \cite{Yaacoby2021}. Analysis at different sizes (Fig. \ref{fig:ph_diag_ws}e) shows a polynomial convergence of this $T_b^*$ towards an asymptotic value $T_b^*\sim2.91$. This suggests that the ME survives the thermodynamic limit and that $N=50$ is a large enough system to capture some properties in the thermodynamic limit of the model. 

%Our analysis highlights a connection between the ME with the phenomenon of \emph{eigenvalue crossing} of the transition matrix $\mathbf{W}$ with respect to the bath temperature $T_b$. In weakly coupled systems an $H=0$ analysis shows how the bath temperature above which the ME can be observed is determined by a crossing of the 2nd and 3rd eigenvalues. Such crossing induces a swap of the corresponding eigenvectors regulating the direction of the slowest relaxation, creating the conditions that allow to observe the ME. In Fig. \ref{fig:ph_diag_ws}(b,d) we plot the relative difference $\delta\lambda_{23}=(\lambda_3 -\lambda_2)/\lambda_2$ as a function of $T_b$, which exhibits a marked dip (indicating the crossing) right before the ME starts to be observed. The crossing naturally defines a "critical" temperature $T_b^*$ that ensures observation of the ME for an antiferromagnet in the region $H<2$.

{Finally,} let us explain the counter-intuitive survival of the Mpmeba effect in the weak coupling limit. The self-thermalization matrix $\mathbf{W}^{ST}$ defined in Eq. \ref{eq:W_ST} allows only transitions within an energy shell, since the bulk is assumed to be isolated from the thermal bath. Transitions between energy shells are realized through $\mathbf{W}^{BC}$ only. However, in the thermodynamic limit, energy shells that have a microscopic energy difference (namely $\Delta E\sim 1$ even though $E\sim N$), might be separated by many boundary, thrusting a slow relaxation among them.
{As a result, the self thermalization in the thermodynamic limit has the same characteristic timescale as that of the boundary transitions.}
It is thus the ideal thermal isolation that is responsible for the far from equilibrium trajectories in the weak coupling limit (Fig. \ref{fig:relax_path}) \footnote{A more general class of anomalous relaxation phenomenon thoroughly analyzed in Ref. \cite{Teza2021}}.

%\textcolor{red}{Therefore, two microstates with the same asymptotic energy shell are not coarse-grained to the same macro-state in the thermodynamic limit, and as a result the self thermalization within each energy shell has the same characteristic time as the boundary transitions.}

%The boundary coupling matrix $\mathbf{W}^{BC}$ (Eq. \ref{eq:W_BC}) connects every energy shell with (at most) 6 other energy shells on a 2-dimensional lattice.The antiferromagnetic character of the system is responsible for the dimensionality: flipping a single spin can either maintain, reduce or increase the chain's domain walls, allowing the system to undertake rel axation shortcuts.

Summarizing, we constructed a theoretical framework to characterize the evolution of a system coupled to the thermal bath only through its boundaries, and proved how anomalous relaxation phenomena can survive such an apparently strong limitation.
{The proposed modelization is quite general, and in principle can be applied to any memoryless system exchanging heath with the thermal bath through limited degrees of freedom.}
Our results corroborate the validity of the ME as an out-of-equilibrium phenomena, proving that it is not an artifact induced by a full coupling. Rather, it seems that it is related to the frustration-like microscopic interactions associated with antiferromagnets. The far-from-equilibrium relaxations in the weak coupling limit are yet another counter-intuitive result, not necessarily related with the ME but linked with the ideal isolation of the bulk of the system \cite{Teza2021}. %The limitations introduced here elevate the description to a faithful representation of large-sized experimental setups in which the ME has been observed \cite{Jeng2006,Chaddah2010,Hu2018,paper:hydrates}, further closing the gap with real-world scenarios. 
The possibility of tuning the coupling strength through $C$, together with the thermodynamics limit analysis (Fig. \ref{fig:ph_diag_ws}) clearly point towards the possibility of experimentally observing the inverse effect in magnetic alloys setups as those utilized in Ref. \cite{Chaddah2010}.

\begin{acknowledgments}
O. R. is the incumbent of the Shlomo and Michla Tomarin career development chair, and is supported by the Abramson Family Center for Young Scientists, the Israel Science Foundation Grant No. 950/19 and by the Minerva foundation.
G. T. is supported by the Center for Statistical Mechanics at the Weizmann Institute of Science, the grant 662962 of the Simons foundation, the grant 873028 of the EU Horizon 2020 program and the NSF-BSF grant 2020765.
We thank David Mukamel and Attilio L. Stella for useful discussions.
\end{acknowledgments}

%S\nocite{*}

\bibliography{refs.bib}% Produces the bibliography via BibTeX.

\providecommand{\noopsort}[1]{}\providecommand{\singleletter}[1]{#1}%
\begin{thebibliography}{41}%
\makeatletter
\providecommand \@ifxundefined [1]{%
 \@ifx{#1\undefined}
}%
\providecommand \@ifnum [1]{%
 \ifnum #1\expandafter \@firstoftwo
 \else \expandafter \@secondoftwo
 \fi
}%
\providecommand \@ifx [1]{%
 \ifx #1\expandafter \@firstoftwo
 \else \expandafter \@secondoftwo
 \fi
}%
\providecommand \natexlab [1]{#1}%
\providecommand \enquote  [1]{``#1''}%
\providecommand \bibnamefont  [1]{#1}%
\providecommand \bibfnamefont [1]{#1}%
\providecommand \citenamefont [1]{#1}%
\providecommand \href@noop [0]{\@secondoftwo}%
\providecommand \href [0]{\begingroup \@sanitize@url \@href}%
\providecommand \@href[1]{\@@startlink{#1}\@@href}%
\providecommand \@@href[1]{\endgroup#1\@@endlink}%
\providecommand \@sanitize@url [0]{\catcode `\\12\catcode `\$12\catcode
  `\&12\catcode `\#12\catcode `\^12\catcode `\_12\catcode `\%12\relax}%
\providecommand \@@startlink[1]{}%
\providecommand \@@endlink[0]{}%
\providecommand \url  [0]{\begingroup\@sanitize@url \@url }%
\providecommand \@url [1]{\endgroup\@href {#1}{\urlprefix }}%
\providecommand \urlprefix  [0]{URL }%
\providecommand \Eprint [0]{\href }%
\providecommand \doibase [0]{https://doi.org/}%
\providecommand \selectlanguage [0]{\@gobble}%
\providecommand \bibinfo  [0]{\@secondoftwo}%
\providecommand \bibfield  [0]{\@secondoftwo}%
\providecommand \translation [1]{[#1]}%
\providecommand \BibitemOpen [0]{}%
\providecommand \bibitemStop [0]{}%
\providecommand \bibitemNoStop [0]{.\EOS\space}%
\providecommand \EOS [0]{\spacefactor3000\relax}%
\providecommand \BibitemShut  [1]{\csname bibitem#1\endcsname}%
\let\auto@bib@innerbib\@empty
%</preamble>
\bibitem [{\citenamefont {Lapolla}\ and\ \citenamefont
  {Godec}(2020)}]{lapolla2020faster}%
  \BibitemOpen
  \bibfield  {author} {\bibinfo {author} {\bibfnamefont {A.}~\bibnamefont
  {Lapolla}}\ and\ \bibinfo {author} {\bibfnamefont {A.}~\bibnamefont
  {Godec}},\ }\bibfield  {title} {\bibinfo {title} {Faster uphill relaxation in
  thermodynamically equidistant temperature quenches},\ }\href@noop {}
  {\bibfield  {journal} {\bibinfo  {journal} {Physical Review Letters}\
  }\textbf {\bibinfo {volume} {125}},\ \bibinfo {pages} {110602} (\bibinfo
  {year} {2020})}\BibitemShut {NoStop}%
\bibitem [{\citenamefont {Gal}\ and\ \citenamefont {Raz}(2020)}]{Gal2020}%
  \BibitemOpen
  \bibfield  {author} {\bibinfo {author} {\bibfnamefont {A.}~\bibnamefont
  {Gal}}\ and\ \bibinfo {author} {\bibfnamefont {O.}~\bibnamefont {Raz}},\
  }\bibfield  {title} {\bibinfo {title} {{Precooling Strategy Allows
  Exponentially Faster Heating}},\ }\bibfield  {journal} {\bibinfo  {journal}
  {Physical Review Letters}\ }\href
  {https://doi.org/10.1103/PhysRevLett.124.060602}
  {10.1103/PhysRevLett.124.060602} (\bibinfo {year} {2020})\BibitemShut
  {NoStop}%
\bibitem [{\citenamefont {Lu}\ and\ \citenamefont {Raz}(2017)}]{Lu2017}%
  \BibitemOpen
  \bibfield  {author} {\bibinfo {author} {\bibfnamefont {Z.}~\bibnamefont
  {Lu}}\ and\ \bibinfo {author} {\bibfnamefont {O.}~\bibnamefont {Raz}},\
  }\bibfield  {title} {\bibinfo {title} {{Nonequilibrium thermodynamics of the
  Markovian Mpemba effect and its inverse}},\ }\bibfield  {journal} {\bibinfo
  {journal} {Proceedings of the National Academy of Sciences of the United
  States of America}\ }\href {https://doi.org/10.1073/pnas.1701264114}
  {10.1073/pnas.1701264114} (\bibinfo {year} {2017})\BibitemShut {NoStop}%
\bibitem [{\citenamefont {Aristotle}()}]{Aristotele350}%
  \BibitemOpen
  \bibfield  {author} {\bibinfo {author} {\bibnamefont {Aristotle}},\
  }\href@noop {} {\emph {\bibinfo {title} {{Meteorology}}}}\ (\bibinfo
  {publisher} {Harvard University Press})\BibitemShut {NoStop}%
\bibitem [{\citenamefont {Mpemba}\ and\ \citenamefont
  {Osborne}(1969)}]{Mpemba1969}%
  \BibitemOpen
  \bibfield  {author} {\bibinfo {author} {\bibfnamefont {E.~B.}\ \bibnamefont
  {Mpemba}}\ and\ \bibinfo {author} {\bibfnamefont {D.~G.}\ \bibnamefont
  {Osborne}},\ }\bibfield  {title} {\bibinfo {title} {{Cool?}},\ }\href@noop {}
  {\bibfield  {journal} {\bibinfo  {journal} {Physics Education}\ }\textbf
  {\bibinfo {volume} {4}},\ \bibinfo {pages} {172} (\bibinfo {year}
  {1969})}\BibitemShut {NoStop}%
\bibitem [{\citenamefont {Jeng}(2006)}]{Jeng2006}%
  \BibitemOpen
  \bibfield  {author} {\bibinfo {author} {\bibfnamefont {M.}~\bibnamefont
  {Jeng}},\ }\bibfield  {title} {\bibinfo {title} {{The Mpemba effect: When can
  hot water freeze faster than cold?}},\ }\bibfield  {journal} {\bibinfo
  {journal} {American Journal of Physics}\ }\href
  {https://doi.org/10.1119/1.2186331} {10.1119/1.2186331} (\bibinfo {year}
  {2006})\BibitemShut {NoStop}%
\bibitem [{\citenamefont {Chaddah}\ \emph {et~al.}(2010)\citenamefont
  {Chaddah}, \citenamefont {Dash}, \citenamefont {Kumar},\ and\ \citenamefont
  {Banerjee}}]{Chaddah2010}%
  \BibitemOpen
  \bibfield  {author} {\bibinfo {author} {\bibfnamefont {P.}~\bibnamefont
  {Chaddah}}, \bibinfo {author} {\bibfnamefont {S.}~\bibnamefont {Dash}},
  \bibinfo {author} {\bibfnamefont {K.}~\bibnamefont {Kumar}},\ and\ \bibinfo
  {author} {\bibfnamefont {A.}~\bibnamefont {Banerjee}},\ }\href
  {https://arxiv.org/abs/1011.3598v1} {\bibinfo {title} {{Overtaking while
  approaching equilibrium}}} (\bibinfo {year} {2010}),\ \Eprint
  {https://arxiv.org/abs/1011.3598} {arXiv:1011.3598} \BibitemShut {NoStop}%
\bibitem [{\citenamefont {Hu}\ \emph {et~al.}(2018)\citenamefont {Hu},
  \citenamefont {Li}, \citenamefont {Huang}, \citenamefont {Li}, \citenamefont
  {Luo}, \citenamefont {Chen}, \citenamefont {Jiang},\ and\ \citenamefont
  {An}}]{Hu2018}%
  \BibitemOpen
  \bibfield  {author} {\bibinfo {author} {\bibfnamefont {C.}~\bibnamefont
  {Hu}}, \bibinfo {author} {\bibfnamefont {J.}~\bibnamefont {Li}}, \bibinfo
  {author} {\bibfnamefont {S.}~\bibnamefont {Huang}}, \bibinfo {author}
  {\bibfnamefont {H.}~\bibnamefont {Li}}, \bibinfo {author} {\bibfnamefont
  {C.}~\bibnamefont {Luo}}, \bibinfo {author} {\bibfnamefont {J.}~\bibnamefont
  {Chen}}, \bibinfo {author} {\bibfnamefont {S.}~\bibnamefont {Jiang}},\ and\
  \bibinfo {author} {\bibfnamefont {L.}~\bibnamefont {An}},\ }\bibfield
  {title} {\bibinfo {title} {{Conformation directed mpeMba effect on
  polylactide crystallization}},\ }\bibfield  {journal} {\bibinfo  {journal}
  {Crystal Growth and Design}\ }\href {https://doi.org/10.1021/acs.cgd.8b01250}
  {10.1021/acs.cgd.8b01250} (\bibinfo {year} {2018})\BibitemShut {NoStop}%
\bibitem [{\citenamefont {Ahn}\ \emph {et~al.}(2016)\citenamefont {Ahn},
  \citenamefont {Kang}, \citenamefont {Koh},\ and\ \citenamefont
  {Lee}}]{paper:hydrates}%
  \BibitemOpen
  \bibfield  {author} {\bibinfo {author} {\bibfnamefont {Y.-H.}\ \bibnamefont
  {Ahn}}, \bibinfo {author} {\bibfnamefont {H.}~\bibnamefont {Kang}}, \bibinfo
  {author} {\bibfnamefont {D.-Y.}\ \bibnamefont {Koh}},\ and\ \bibinfo {author}
  {\bibfnamefont {H.}~\bibnamefont {Lee}},\ }\bibfield  {title} {\bibinfo
  {title} {Experimental verifications of mpemba-like behaviors of clathrate
  hydrates},\ }\href@noop {} {\bibfield  {journal} {\bibinfo  {journal} {Korean
  Journal of Chemical Engineering}\ ,\ \bibinfo {pages} {1}} (\bibinfo {year}
  {2016})}\BibitemShut {NoStop}%
\bibitem [{\citenamefont {Kumar}\ and\ \citenamefont
  {Bechhoefer}(2020)}]{Kumar2020}%
  \BibitemOpen
  \bibfield  {author} {\bibinfo {author} {\bibfnamefont {A.}~\bibnamefont
  {Kumar}}\ and\ \bibinfo {author} {\bibfnamefont {J.}~\bibnamefont
  {Bechhoefer}},\ }\bibfield  {title} {\bibinfo {title} {{Exponentially faster
  cooling in a colloidal system}},\ }\bibfield  {journal} {\bibinfo  {journal}
  {Nature}\ }\href {https://doi.org/10.1038/s41586-020-2560-x}
  {10.1038/s41586-020-2560-x} (\bibinfo {year} {2020}),\ \Eprint
  {https://arxiv.org/abs/2008.02373} {arXiv:2008.02373} \BibitemShut {NoStop}%
\bibitem [{\citenamefont {Kumar}\ \emph {et~al.}(2021)\citenamefont {Kumar},
  \citenamefont {Chetrite},\ and\ \citenamefont
  {Bechhoefer}}]{kumar2021anomalous}%
  \BibitemOpen
  \bibfield  {author} {\bibinfo {author} {\bibfnamefont {A.}~\bibnamefont
  {Kumar}}, \bibinfo {author} {\bibfnamefont {R.}~\bibnamefont {Chetrite}},\
  and\ \bibinfo {author} {\bibfnamefont {J.}~\bibnamefont {Bechhoefer}},\
  }\bibfield  {title} {\bibinfo {title} {Anomalous heating in a colloidal
  system},\ }\href@noop {} {\bibfield  {journal} {\bibinfo  {journal} {arXiv
  preprint arXiv:2104.12899}\ } (\bibinfo {year} {2021})}\BibitemShut {NoStop}%
\bibitem [{\citenamefont {Tao}\ \emph {et~al.}(2016)\citenamefont {Tao},
  \citenamefont {Zou}, \citenamefont {Jia}, \citenamefont {Li},\ and\
  \citenamefont {Cremer}}]{Tao2016}%
  \BibitemOpen
  \bibfield  {author} {\bibinfo {author} {\bibfnamefont {Y.}~\bibnamefont
  {Tao}}, \bibinfo {author} {\bibfnamefont {W.}~\bibnamefont {Zou}}, \bibinfo
  {author} {\bibfnamefont {J.}~\bibnamefont {Jia}}, \bibinfo {author}
  {\bibfnamefont {W.}~\bibnamefont {Li}},\ and\ \bibinfo {author}
  {\bibfnamefont {D.}~\bibnamefont {Cremer}},\ }\bibfield  {title} {\bibinfo
  {title} {{Different Ways of Hydrogen Bonding in Water - Why Does Warm Water
  Freeze Faster than Cold Water?}},\ }\href
  {https://doi.org/10.1021/ACS.JCTC.6B00735} {\bibfield  {journal} {\bibinfo
  {journal} {Journal of Chemical Theory and Computation}\ }\textbf {\bibinfo
  {volume} {13}},\ \bibinfo {pages} {55} (\bibinfo {year} {2016})}\BibitemShut
  {NoStop}%
\bibitem [{\citenamefont {Zhang}\ \emph {et~al.}(2014)\citenamefont {Zhang},
  \citenamefont {Huang}, \citenamefont {Ma}, \citenamefont {Zhou},
  \citenamefont {Zhou}, \citenamefont {Zheng}, \citenamefont {Jiang},\ and\
  \citenamefont {Sun}}]{Zhang2014}%
  \BibitemOpen
  \bibfield  {author} {\bibinfo {author} {\bibfnamefont {X.}~\bibnamefont
  {Zhang}}, \bibinfo {author} {\bibfnamefont {Y.}~\bibnamefont {Huang}},
  \bibinfo {author} {\bibfnamefont {Z.}~\bibnamefont {Ma}}, \bibinfo {author}
  {\bibfnamefont {Y.}~\bibnamefont {Zhou}}, \bibinfo {author} {\bibfnamefont
  {J.}~\bibnamefont {Zhou}}, \bibinfo {author} {\bibfnamefont {W.}~\bibnamefont
  {Zheng}}, \bibinfo {author} {\bibfnamefont {Q.}~\bibnamefont {Jiang}},\ and\
  \bibinfo {author} {\bibfnamefont {C.~Q.}\ \bibnamefont {Sun}},\ }\bibfield
  {title} {\bibinfo {title} {{Hydrogen-bond memory and water-skin supersolidity
  resolving the Mpemba paradox}},\ }\href {https://doi.org/10.1039/C4CP03669G}
  {\bibfield  {journal} {\bibinfo  {journal} {Physical Chemistry Chemical
  Physics}\ }\textbf {\bibinfo {volume} {16}},\ \bibinfo {pages} {22995}
  (\bibinfo {year} {2014})}\BibitemShut {NoStop}%
\bibitem [{\citenamefont {Vynnycky}\ and\ \citenamefont
  {Kimura}(2015)}]{Vynnycky2015}%
  \BibitemOpen
  \bibfield  {author} {\bibinfo {author} {\bibfnamefont {M.}~\bibnamefont
  {Vynnycky}}\ and\ \bibinfo {author} {\bibfnamefont {S.}~\bibnamefont
  {Kimura}},\ }\bibfield  {title} {\bibinfo {title} {{Can natural convection
  alone explain the Mpemba effect?}},\ }\href
  {https://doi.org/10.1016/J.IJHEATMASSTRANSFER.2014.09.015} {\bibfield
  {journal} {\bibinfo  {journal} {International Journal of Heat and Mass
  Transfer}\ }\textbf {\bibinfo {volume} {80}},\ \bibinfo {pages} {243}
  (\bibinfo {year} {2015})}\BibitemShut {NoStop}%
\bibitem [{\citenamefont {Auerbach}(1998)}]{Auerbach1998}%
  \BibitemOpen
  \bibfield  {author} {\bibinfo {author} {\bibfnamefont {D.}~\bibnamefont
  {Auerbach}},\ }\bibfield  {title} {\bibinfo {title} {{Supercooling and the
  Mpemba effect: When hot water freezes quicker than cold}},\ }\href
  {https://doi.org/10.1119/1.18059} {\bibfield  {journal} {\bibinfo  {journal}
  {American Journal of Physics}\ }\textbf {\bibinfo {volume} {63}},\ \bibinfo
  {pages} {882} (\bibinfo {year} {1998})}\BibitemShut {NoStop}%
\bibitem [{\citenamefont {Mirabedin}\ and\ \citenamefont
  {Farhadi}(2017)}]{Mirabedin2017}%
  \BibitemOpen
  \bibfield  {author} {\bibinfo {author} {\bibfnamefont {S.~M.}\ \bibnamefont
  {Mirabedin}}\ and\ \bibinfo {author} {\bibfnamefont {F.}~\bibnamefont
  {Farhadi}},\ }\bibfield  {title} {\bibinfo {title} {{Numerical investigation
  of solidification of single droplets with and without evaporation
  mechanism}},\ }\href {https://doi.org/10.1016/J.IJREFRIG.2016.09.006}
  {\bibfield  {journal} {\bibinfo  {journal} {International Journal of
  Refrigeration}\ }\textbf {\bibinfo {volume} {73}},\ \bibinfo {pages} {219}
  (\bibinfo {year} {2017})}\BibitemShut {NoStop}%
\bibitem [{\citenamefont {Lasanta}\ \emph {et~al.}(2017)\citenamefont
  {Lasanta}, \citenamefont {{Vega Reyes}}, \citenamefont {Prados},\ and\
  \citenamefont {Santos}}]{Lasanta2017}%
  \BibitemOpen
  \bibfield  {author} {\bibinfo {author} {\bibfnamefont {A.}~\bibnamefont
  {Lasanta}}, \bibinfo {author} {\bibfnamefont {F.}~\bibnamefont {{Vega
  Reyes}}}, \bibinfo {author} {\bibfnamefont {A.}~\bibnamefont {Prados}},\ and\
  \bibinfo {author} {\bibfnamefont {A.}~\bibnamefont {Santos}},\ }\bibfield
  {title} {\bibinfo {title} {{When the Hotter Cools More Quickly: Mpemba Effect
  in Granular Fluids}},\ }\bibfield  {journal} {\bibinfo  {journal} {Physical
  Review Letters}\ }\href {https://doi.org/10.1103/PhysRevLett.119.148001}
  {10.1103/PhysRevLett.119.148001} (\bibinfo {year} {2017}),\ \Eprint
  {https://arxiv.org/abs/1611.04948} {arXiv:1611.04948} \BibitemShut {NoStop}%
\bibitem [{\citenamefont {Torrente}\ \emph {et~al.}(2019)\citenamefont
  {Torrente}, \citenamefont {L{\'o}pez-Casta{\~n}o}, \citenamefont {Lasanta},
  \citenamefont {Reyes}, \citenamefont {Prados},\ and\ \citenamefont
  {Santos}}]{torrente2019large}%
  \BibitemOpen
  \bibfield  {author} {\bibinfo {author} {\bibfnamefont {A.}~\bibnamefont
  {Torrente}}, \bibinfo {author} {\bibfnamefont {M.~A.}\ \bibnamefont
  {L{\'o}pez-Casta{\~n}o}}, \bibinfo {author} {\bibfnamefont {A.}~\bibnamefont
  {Lasanta}}, \bibinfo {author} {\bibfnamefont {F.~V.}\ \bibnamefont {Reyes}},
  \bibinfo {author} {\bibfnamefont {A.}~\bibnamefont {Prados}},\ and\ \bibinfo
  {author} {\bibfnamefont {A.}~\bibnamefont {Santos}},\ }\bibfield  {title}
  {\bibinfo {title} {Large mpemba-like effect in a gas of inelastic rough hard
  spheres},\ }\href@noop {} {\bibfield  {journal} {\bibinfo  {journal}
  {Physical Review E}\ }\textbf {\bibinfo {volume} {99}},\ \bibinfo {pages}
  {060901} (\bibinfo {year} {2019})}\BibitemShut {NoStop}%
\bibitem [{\citenamefont {Biswas}\ \emph {et~al.}(2020)\citenamefont {Biswas},
  \citenamefont {Prasad}, \citenamefont {Raz},\ and\ \citenamefont
  {Rajesh}}]{biswas2020mpemba}%
  \BibitemOpen
  \bibfield  {author} {\bibinfo {author} {\bibfnamefont {A.}~\bibnamefont
  {Biswas}}, \bibinfo {author} {\bibfnamefont {V.}~\bibnamefont {Prasad}},
  \bibinfo {author} {\bibfnamefont {O.}~\bibnamefont {Raz}},\ and\ \bibinfo
  {author} {\bibfnamefont {R.}~\bibnamefont {Rajesh}},\ }\bibfield  {title}
  {\bibinfo {title} {Mpemba effect in driven granular maxwell gases},\
  }\href@noop {} {\bibfield  {journal} {\bibinfo  {journal} {Physical Review
  E}\ }\textbf {\bibinfo {volume} {102}},\ \bibinfo {pages} {012906} (\bibinfo
  {year} {2020})}\BibitemShut {NoStop}%
\bibitem [{\citenamefont {Santos}\ and\ \citenamefont
  {Prados}(2020)}]{santos2020mpemba}%
  \BibitemOpen
  \bibfield  {author} {\bibinfo {author} {\bibfnamefont {A.}~\bibnamefont
  {Santos}}\ and\ \bibinfo {author} {\bibfnamefont {A.}~\bibnamefont
  {Prados}},\ }\bibfield  {title} {\bibinfo {title} {Mpemba effect in molecular
  gases under nonlinear drag},\ }\href@noop {} {\bibfield  {journal} {\bibinfo
  {journal} {Physics of Fluids}\ }\textbf {\bibinfo {volume} {32}},\ \bibinfo
  {pages} {072010} (\bibinfo {year} {2020})}\BibitemShut {NoStop}%
\bibitem [{\citenamefont {Takada}\ \emph {et~al.}(2021)\citenamefont {Takada},
  \citenamefont {Hayakawa},\ and\ \citenamefont {Santos}}]{takada2021mpemba}%
  \BibitemOpen
  \bibfield  {author} {\bibinfo {author} {\bibfnamefont {S.}~\bibnamefont
  {Takada}}, \bibinfo {author} {\bibfnamefont {H.}~\bibnamefont {Hayakawa}},\
  and\ \bibinfo {author} {\bibfnamefont {A.}~\bibnamefont {Santos}},\
  }\bibfield  {title} {\bibinfo {title} {Mpemba effect in inertial
  suspensions},\ }\href@noop {} {\bibfield  {journal} {\bibinfo  {journal}
  {Physical Review E}\ }\textbf {\bibinfo {volume} {103}},\ \bibinfo {pages}
  {032901} (\bibinfo {year} {2021})}\BibitemShut {NoStop}%
\bibitem [{\citenamefont {Momp{\'o}}\ \emph {et~al.}(2021)\citenamefont
  {Momp{\'o}}, \citenamefont {L{\'o}pez-Casta{\~n}o}, \citenamefont {Lasanta},
  \citenamefont {Vega~Reyes},\ and\ \citenamefont
  {Torrente}}]{mompo2021memory}%
  \BibitemOpen
  \bibfield  {author} {\bibinfo {author} {\bibfnamefont {E.}~\bibnamefont
  {Momp{\'o}}}, \bibinfo {author} {\bibfnamefont {M.}~\bibnamefont
  {L{\'o}pez-Casta{\~n}o}}, \bibinfo {author} {\bibfnamefont {A.}~\bibnamefont
  {Lasanta}}, \bibinfo {author} {\bibfnamefont {F.}~\bibnamefont
  {Vega~Reyes}},\ and\ \bibinfo {author} {\bibfnamefont {A.}~\bibnamefont
  {Torrente}},\ }\bibfield  {title} {\bibinfo {title} {Memory effects in a gas
  of viscoelastic particles},\ }\href@noop {} {\bibfield  {journal} {\bibinfo
  {journal} {Physics of Fluids}\ }\textbf {\bibinfo {volume} {33}},\ \bibinfo
  {pages} {062005} (\bibinfo {year} {2021})}\BibitemShut {NoStop}%
\bibitem [{\citenamefont {Ch{\'e}trite}\ \emph {et~al.}(2021)\citenamefont
  {Ch{\'e}trite}, \citenamefont {Kumar},\ and\ \citenamefont
  {Bechhoefer}}]{chetrite2021metastable}%
  \BibitemOpen
  \bibfield  {author} {\bibinfo {author} {\bibfnamefont {R.}~\bibnamefont
  {Ch{\'e}trite}}, \bibinfo {author} {\bibfnamefont {A.}~\bibnamefont
  {Kumar}},\ and\ \bibinfo {author} {\bibfnamefont {J.}~\bibnamefont
  {Bechhoefer}},\ }\bibfield  {title} {\bibinfo {title} {The metastable mpemba
  effect corresponds to a non-monotonic temperature dependence of extractable
  work},\ }\href@noop {} {\bibfield  {journal} {\bibinfo  {journal} {Frontiers
  in Physics}\ }\textbf {\bibinfo {volume} {9}},\ \bibinfo {pages} {141}
  (\bibinfo {year} {2021})}\BibitemShut {NoStop}%
\bibitem [{\citenamefont {Walker}\ and\ \citenamefont
  {Vucelja}(2021)}]{walker2021anomalous}%
  \BibitemOpen
  \bibfield  {author} {\bibinfo {author} {\bibfnamefont {M.}~\bibnamefont
  {Walker}}\ and\ \bibinfo {author} {\bibfnamefont {M.}~\bibnamefont
  {Vucelja}},\ }\bibfield  {title} {\bibinfo {title} {Anomalous thermal
  relaxation of langevin particles in a piecewise constant potential},\
  }\href@noop {} {\bibfield  {journal} {\bibinfo  {journal} {arXiv preprint
  arXiv:2105.10656}\ } (\bibinfo {year} {2021})}\BibitemShut {NoStop}%
\bibitem [{\citenamefont {Busiello}\ \emph {et~al.}(2021)\citenamefont
  {Busiello}, \citenamefont {Gupta},\ and\ \citenamefont
  {Maritan}}]{busiello2021inducing}%
  \BibitemOpen
  \bibfield  {author} {\bibinfo {author} {\bibfnamefont {D.~M.}\ \bibnamefont
  {Busiello}}, \bibinfo {author} {\bibfnamefont {D.}~\bibnamefont {Gupta}},\
  and\ \bibinfo {author} {\bibfnamefont {A.}~\bibnamefont {Maritan}},\
  }\bibfield  {title} {\bibinfo {title} {Inducing and optimizing markovian
  mpemba effect with stochastic reset},\ }\href@noop {} {\bibfield  {journal}
  {\bibinfo  {journal} {arXiv preprint arXiv:2106.08044}\ } (\bibinfo {year}
  {2021})}\BibitemShut {NoStop}%
\bibitem [{\citenamefont {Carollo}\ \emph {et~al.}(2021)\citenamefont
  {Carollo}, \citenamefont {Lasanta},\ and\ \citenamefont
  {Lesanovsky}}]{Carollo2021}%
  \BibitemOpen
  \bibfield  {author} {\bibinfo {author} {\bibfnamefont {F.}~\bibnamefont
  {Carollo}}, \bibinfo {author} {\bibfnamefont {A.}~\bibnamefont {Lasanta}},\
  and\ \bibinfo {author} {\bibfnamefont {I.}~\bibnamefont {Lesanovsky}},\
  }\bibfield  {title} {\bibinfo {title} {{Exponentially Accelerated Approach to
  Stationarity in Markovian Open Quantum Systems through the Mpemba Effect}},\
  }\href {https://doi.org/10.1103/PhysRevLett.127.060401} {\bibfield  {journal}
  {\bibinfo  {journal} {Physical Review Letters}\ }\textbf {\bibinfo {volume}
  {127}},\ \bibinfo {pages} {060401} (\bibinfo {year} {2021})}\BibitemShut
  {NoStop}%
\bibitem [{\citenamefont {Baity-Jesi}\ \emph {et~al.}(2019)\citenamefont
  {Baity-Jesi}, \citenamefont {Calore}, \citenamefont {Cruz}, \citenamefont
  {Fernandez}, \citenamefont {Gil-Narvi{\'o}n}, \citenamefont
  {Gordillo-Guerrero}, \citenamefont {I{\~n}iguez}, \citenamefont {Lasanta},
  \citenamefont {Maiorano}, \citenamefont {Marinari} \emph
  {et~al.}}]{baity2019mpemba}%
  \BibitemOpen
  \bibfield  {author} {\bibinfo {author} {\bibfnamefont {M.}~\bibnamefont
  {Baity-Jesi}}, \bibinfo {author} {\bibfnamefont {E.}~\bibnamefont {Calore}},
  \bibinfo {author} {\bibfnamefont {A.}~\bibnamefont {Cruz}}, \bibinfo {author}
  {\bibfnamefont {L.~A.}\ \bibnamefont {Fernandez}}, \bibinfo {author}
  {\bibfnamefont {J.~M.}\ \bibnamefont {Gil-Narvi{\'o}n}}, \bibinfo {author}
  {\bibfnamefont {A.}~\bibnamefont {Gordillo-Guerrero}}, \bibinfo {author}
  {\bibfnamefont {D.}~\bibnamefont {I{\~n}iguez}}, \bibinfo {author}
  {\bibfnamefont {A.}~\bibnamefont {Lasanta}}, \bibinfo {author} {\bibfnamefont
  {A.}~\bibnamefont {Maiorano}}, \bibinfo {author} {\bibfnamefont
  {E.}~\bibnamefont {Marinari}}, \emph {et~al.},\ }\bibfield  {title} {\bibinfo
  {title} {The mpemba effect in spin glasses is a persistent memory effect},\
  }\href@noop {} {\bibfield  {journal} {\bibinfo  {journal} {Proceedings of the
  National Academy of Sciences}\ }\textbf {\bibinfo {volume} {116}},\ \bibinfo
  {pages} {15350} (\bibinfo {year} {2019})}\BibitemShut {NoStop}%
\bibitem [{\citenamefont {Nava}\ and\ \citenamefont
  {Fabrizio}(2019)}]{nava2019lindblad}%
  \BibitemOpen
  \bibfield  {author} {\bibinfo {author} {\bibfnamefont {A.}~\bibnamefont
  {Nava}}\ and\ \bibinfo {author} {\bibfnamefont {M.}~\bibnamefont
  {Fabrizio}},\ }\bibfield  {title} {\bibinfo {title} {Lindblad dissipative
  dynamics in the presence of phase coexistence},\ }\href@noop {} {\bibfield
  {journal} {\bibinfo  {journal} {Physical Review B}\ }\textbf {\bibinfo
  {volume} {100}},\ \bibinfo {pages} {125102} (\bibinfo {year}
  {2019})}\BibitemShut {NoStop}%
\bibitem [{\citenamefont {Yang}\ and\ \citenamefont {Hou}(2020)}]{yang2020non}%
  \BibitemOpen
  \bibfield  {author} {\bibinfo {author} {\bibfnamefont {Z.-Y.}\ \bibnamefont
  {Yang}}\ and\ \bibinfo {author} {\bibfnamefont {J.-X.}\ \bibnamefont {Hou}},\
  }\bibfield  {title} {\bibinfo {title} {Non-markovian mpemba effect in
  mean-field systems},\ }\href@noop {} {\bibfield  {journal} {\bibinfo
  {journal} {Physical Review E}\ }\textbf {\bibinfo {volume} {101}},\ \bibinfo
  {pages} {052106} (\bibinfo {year} {2020})}\BibitemShut {NoStop}%
\bibitem [{\citenamefont {Vadakkayil}\ and\ \citenamefont
  {Das}(2021)}]{vadakkayil2021should}%
  \BibitemOpen
  \bibfield  {author} {\bibinfo {author} {\bibfnamefont {N.}~\bibnamefont
  {Vadakkayil}}\ and\ \bibinfo {author} {\bibfnamefont {S.~K.}\ \bibnamefont
  {Das}},\ }\bibfield  {title} {\bibinfo {title} {Should a hotter paramagnet
  transform quicker to a ferromagnet? monte carlo simulation results for ising
  model},\ }\href@noop {} {\bibfield  {journal} {\bibinfo  {journal} {Physical
  Chemistry Chemical Physics}\ }\textbf {\bibinfo {volume} {23}},\ \bibinfo
  {pages} {11186} (\bibinfo {year} {2021})}\BibitemShut {NoStop}%
\bibitem [{\citenamefont {Klich}\ \emph {et~al.}(2019)\citenamefont {Klich},
  \citenamefont {Raz}, \citenamefont {Hirschberg},\ and\ \citenamefont
  {Vucelja}}]{Klich2019}%
  \BibitemOpen
  \bibfield  {author} {\bibinfo {author} {\bibfnamefont {I.}~\bibnamefont
  {Klich}}, \bibinfo {author} {\bibfnamefont {O.}~\bibnamefont {Raz}}, \bibinfo
  {author} {\bibfnamefont {O.}~\bibnamefont {Hirschberg}},\ and\ \bibinfo
  {author} {\bibfnamefont {M.}~\bibnamefont {Vucelja}},\ }\bibfield  {title}
  {\bibinfo {title} {{Mpemba Index and Anomalous Relaxation}},\ }\bibfield
  {journal} {\bibinfo  {journal} {Physical Review X}\ }\href
  {https://doi.org/10.1103/PhysRevX.9.021060} {10.1103/PhysRevX.9.021060}
  (\bibinfo {year} {2019}),\ \Eprint {https://arxiv.org/abs/1711.05829}
  {arXiv:1711.05829} \BibitemShut {NoStop}%
\bibitem [{\citenamefont {Teza}\ and\ \citenamefont {Stella}(2020)}]{Teza2020}%
  \BibitemOpen
  \bibfield  {author} {\bibinfo {author} {\bibfnamefont {G.}~\bibnamefont
  {Teza}}\ and\ \bibinfo {author} {\bibfnamefont {A.~L.}\ \bibnamefont
  {Stella}},\ }\bibfield  {title} {\bibinfo {title} {Exact coarse graining
  preserves entropy production out of equilibrium},\ }\href
  {https://doi.org/10.1103/PhysRevLett.125.110601} {\bibfield  {journal}
  {\bibinfo  {journal} {Phys. Rev. Lett.}\ }\textbf {\bibinfo {volume} {125}},\
  \bibinfo {pages} {110601} (\bibinfo {year} {2020})}\BibitemShut {NoStop}%
\bibitem [{\citenamefont {Teza}(2020)}]{Teza2020b}%
  \BibitemOpen
  \bibfield  {author} {\bibinfo {author} {\bibfnamefont {G.}~\bibnamefont
  {Teza}},\ }\emph {\bibinfo {title} {Out of equilibrium dynamics: from an
  entropy of the growth to the growth of entropy production}},\ \href
  {http://paduaresearch.cab.unipd.it/12995/} {Ph.D. thesis},\ \bibinfo
  {school} {University of Padova} (\bibinfo {year} {2020})\BibitemShut
  {NoStop}%
\bibitem [{Note1()}]{Note1}%
  \BibitemOpen
  \bibinfo {note} {Similar analysis can be made in continuous frameworks as in
  \cite {Lu2017}}\BibitemShut {NoStop}%
\bibitem [{Note2()}]{Note2}%
  \BibitemOpen
  \bibinfo {note} {The specific normalization chosen for $\protect \mathbf
  {W}^{\{BC,ST\}}$ changes the value of $C$, but not its limiting cases.
  Specifically, we chose to normalize with respect to the maximum rate so that
  $\protect \qopname \relax m{max}(\protect \mathbf {W}_{ij}^{ST})=\protect
  \qopname \relax m{max}(\protect \mathbf {W}_{ij}^{BC})\equiv 1$.}\BibitemShut
  {Stop}%
\bibitem [{\citenamefont {Glauber}(1963)}]{Glauber1963}%
  \BibitemOpen
  \bibfield  {author} {\bibinfo {author} {\bibfnamefont {R.~J.}\ \bibnamefont
  {Glauber}},\ }\bibfield  {title} {\bibinfo {title} {{Time-dependent
  statistics of the Ising model}},\ }\href {https://doi.org/10.1063/1.1703954}
  {\bibfield  {journal} {\bibinfo  {journal} {Journal of Mathematical Physics}\
  }\textbf {\bibinfo {volume} {4}},\ \bibinfo {pages} {294} (\bibinfo {year}
  {1963})}\BibitemShut {NoStop}%
\bibitem [{\citenamefont {Felderhof}(1971)}]{Felderhof1971}%
  \BibitemOpen
  \bibfield  {author} {\bibinfo {author} {\bibfnamefont {B.~U.}\ \bibnamefont
  {Felderhof}},\ }\bibfield  {title} {\bibinfo {title} {{Spin relaxation of the
  Ising chain}},\ }\href {https://doi.org/10.1016/S0034-4877(71)80006-X}
  {\bibfield  {journal} {\bibinfo  {journal} {Reports on Mathematical Physics}\
  }\textbf {\bibinfo {volume} {1}},\ \bibinfo {pages} {215} (\bibinfo {year}
  {1971})}\BibitemShut {NoStop}%
\bibitem [{\citenamefont {Hamming}(1950)}]{Hamming1950}%
  \BibitemOpen
  \bibfield  {author} {\bibinfo {author} {\bibfnamefont {R.~W.}\ \bibnamefont
  {Hamming}},\ }\bibfield  {title} {\bibinfo {title} {{Error Detecting and
  Error Correcting Codes}},\ }\href
  {https://doi.org/10.1002/j.1538-7305.1950.tb00463.x} {\bibfield  {journal}
  {\bibinfo  {journal} {Bell System Technical Journal}\ }\textbf {\bibinfo
  {volume} {29}},\ \bibinfo {pages} {147} (\bibinfo {year} {1950})}\BibitemShut
  {NoStop}%
\bibitem [{\citenamefont {Yaacoby}\ \emph {et~al.}(2022)\citenamefont
  {Yaacoby}, \citenamefont {Teza},\ and\ \citenamefont {Raz}}]{Yaacoby2021}%
  \BibitemOpen
  \bibfield  {author} {\bibinfo {author} {\bibfnamefont {R.}~\bibnamefont
  {Yaacoby}}, \bibinfo {author} {\bibfnamefont {G.}~\bibnamefont {Teza}},\ and\
  \bibinfo {author} {\bibfnamefont {O.}~\bibnamefont {Raz}},\ }\href@noop {}
  {\bibinfo {title} {{Eigenvector swapping in Markovian dynamics (in
  preparation)}}} (\bibinfo {year} {2022})\BibitemShut {NoStop}%
\bibitem [{Note3()}]{Note3}%
  \BibitemOpen
  \bibinfo {note} {A more general class of anomalous relaxation phenomenon
  thoroughly analyzed in Ref. \cite {Teza2021}}\BibitemShut {NoStop}%
\bibitem [{\citenamefont {Teza}\ \emph {et~al.}(2022)\citenamefont {Teza},
  \citenamefont {Yaacoby},\ and\ \citenamefont {Raz}}]{Teza2021}%
  \BibitemOpen
  \bibfield  {author} {\bibinfo {author} {\bibfnamefont {G.}~\bibnamefont
  {Teza}}, \bibinfo {author} {\bibfnamefont {R.}~\bibnamefont {Yaacoby}},\ and\
  \bibinfo {author} {\bibfnamefont {O.}~\bibnamefont {Raz}},\ }\href@noop {}
  {\bibinfo {title} {{Out of equilibrium relaxation in weak couplings (in
  preparation)}}} (\bibinfo {year} {2022})\BibitemShut {NoStop}%
\end{thebibliography}%

\end{document}